\documentclass[twocolumn,showpacs,amsmath,amssymb,pr,superscriptaddress]{revtex4-1}
\usepackage{graphicx}
\usepackage{epsf}
\usepackage{ulem}
\usepackage{color}
\usepackage[unicode=true,colorlinks=true,urlcolor=blue,citecolor=blue]{hyperref}

\begin{document}

\title{Nonlinear AC and DC Conductivities
in a Two-Subband n-GaAs/AlAs Heterostructure
}

\author{I. L. Drichko}
\author{I. Yu. Smirnov}
\affiliation{Ioffe Institute, Russian Academy of Sciences, St. Petersburg, 194021 Russia}
\author{A. K. Bakarov}
\author{A. A. Bykov}
\affiliation{Rzhanov Institute of Semiconductor Physics, Siberian Branch, Russian Academy of Sciences, Novosibirsk, 630090 Russia}
\author{A. A. Dmitriev}
\affiliation{ITMO University, St. Petersburg, 197101 Russia}
\author{Yu. M. Galperin}
\affiliation{Department of Physics, University of Oslo, P. O. Box 1048 Blindern, Oslo, 0316 Norway}
\affiliation{Ioffe Institute, Russian Academy of Sciences, St. Petersburg, 194021 Russia}

\begin{abstract}
The DC and AC conductivities of the n-GaAs/AlAs heterostructure with two filled size quantization levels are studied within a wide magnetic field range. The electron spectrum of such heterostructure is characterized by two subbands (symmetric $S$ and antisymmetric $AS$), separated by the band gap $\Delta_{12}=15.5$~meV. It is shown that, in the linear regime at the applied magnetic field $B >3$~T, the system exhibits oscillations corresponding to the integer quantum Hall effect. A quite complicated pattern of such oscillations is well interpreted in terms of transitions between Landau levels related to different subbands. At $B <1$~T, magneto-intersubband resistance oscillations (MISOs) are observed. An increase in the conductivity with the electric current flowing across the sample or in the intensity of the surface acoustic wave (SAW) in the regime of the integer quantum Hall effect is determined by an increase in the electron gas temperature. In the case of intersubband transitions, it is found that nonlinearity cannot be explained by heating. At the same time, the decrease in the AC conductivity with increasing SAW electric field is independent of frequency, but the corresponding behavior does not coincide with that corresponding to the dependence of the DC conductivity on the Hall voltage $E_y$.
\end{abstract}


\maketitle

\section{Introduction}

The electron spectrum of semiconductor heterostructures including two quantum wells, wide quantum wells, or two size quantization bands with bottoms below the Fermi level has two subbands separated by a band gap $\Delta_{12}$. Coupling between these subbands significantly affects the main characteristics of such two-subband systems, giving rise to a number of new magnetotransport phenomena~\cite{Polyanovskii88,Boebinger90}, which are absent in single-subband systems. For example, in a two-subband system, the 1/$B$ dependence of the conductivity exhibits not only periodic Shubnikov–de Haas (SdH) oscillations whose frequencies ($f_1$ and $f_2$) are determined by the electron densities in the subbands ($n_1$ and $n_2$) but also oscillations with a difference frequency ($f_1$ - $f_2$). These oscillations, referred to as magneto-intersubband oscillations (MISOs), are due to transitions between the states corresponding to the same energy occurring when Landau levels cross different subbands. The resonance nature of such interband transitions does not depend on the position of the Fermi level and, therefore, MISOs appear at higher temperatures than those characteristic of the SdH oscillations~\cite{Polyanovskii88}. Magneto-intersubband oscillations were actively studied both theoretically~\cite{Polyanovskii88,Raikh94,Averkiev01,Raichev08} and experimentally in single and double GaAs quantum wells~\cite{Leadley92,Bykov563}. Recently, they were detected in a HgTe quantum well with two filled spin subbands~\cite{Minkov274}. In quantizing magnetic fields, two-subband systems exhibit not only the integer and fractional quantum Hall effects~\cite{Boebinger90,Suen} but also collective electronic states caused by the anticrossing of Landau levels of different subbands~\cite{Lee,Zhang}.

Such heterostructures also exhibit unusual non-ohmic effects arising with an increase in the electric current flowing across the sample under study at low magnetic fields, at which the intersubband transitions are observed~\cite{Bykov08,Mamani09,Wiedmann11,Dietrich12}. Despite the long-term history of the research in the field of two-subband electron systems, many aspects of magnetotransport in them are still debatable ~\cite{DrichkoIntersub,bib:Bykov2,Dmitriev19}. In the presence of two partially filled subbands, the picture of Shubnikov–de Haas oscillations (as well as the picture of the integer Hall effect) is quite complicated and requires special investigation.

In this work, we study the n-GaAs/AlAs heterostructure with the 26-nm-wide potential well and with AlAs/GaAs superlattice potential barriers. The DC transport characteristics of this heterostructure measured both in linear and in nonlinear regimes were studied in detail in~\cite{Dietrich12,Goran09,Bykov81,Mayer,Bykov100} at magnetic fields up to 2~T. These studies demonstrate that the total charge carrier (electron) density $n_{\mathrm{tot}}$ equals $8.13 \times 10^{11}$~cm$^{-2}$; hence, the upper (second) size quantization level turns out to be below the Fermi level. Therefore, the electron spectrum has two subbands (symmetric and antisymmetric) separated by an energy gap $\Delta_{12}=15.5$meV. The charge carrier densities in these subbands differ by a factor of 3: in the symmetric subband, $n_1=6.2\times 10^{11}$~cm$^{-2}$, and in the antisymmetric one, $n_2=1.9 \times 10^{11}$~cm$^{-2}$. These data are obtained by the Fourier analysis of DC conductivity oscillations.

In this work, the effect of the two-subband energy spectrum on the formation of magnetotransport oscillation patterns at applied magnetic fields up to 14 T in the linear and nonlinear regimes is studied using DC measurements (in fields up to 14~T) and contactless acoustic spectroscopy (in fields up to 8~T). As far as we know, such measurements in two-subband structures have not yet been performed. In particular, we are going to study the frequency dependence of the AC conductivity in the nonlinear regime.

\section{Experimental techniques and results}

\begin{figure}[h]
\centerline{
\includegraphics[width=8cm,clip=]{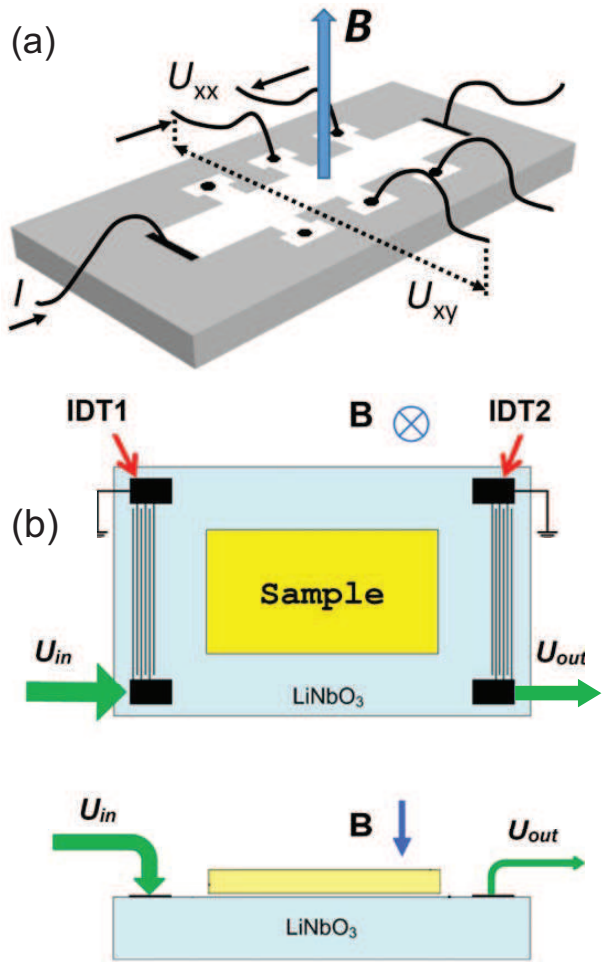}}
\caption{Experimental techniques and actual
parameter ranges: (a) DC measurements (with the Hall
bar). Simultaneous measurements of $\sigma_{xx} (B)$ and $\sigma_{xy} (B)$ at
$B \leq$~14~T, $T=$~2--20~K. (b) Acoustic technique. Determination
of the AC conductivity $\sigma_{xx}(\omega)\equiv \sigma_1-i\sigma_2$. $B \leq$~8~T, $T=$~1.7--15~K.} \label{fig1}
\end{figure}

The used experimental techniques and the actual ranges of the measured characteristics are illustrated in Fig.\ref{fig1}. A more detailed description can be found, e.g., in~\cite{Dmitriev19}.

The DC measurements are performed using a $50\times 450$~$\mu$m$^2$ Hall bar, whereas the $\rho_{xx} (B)$ and $\rho_{xy} (B)$ components of magnetoresistance are studied at magnetic fields up to 14~T and at temperatures from 2.2 to 20~K in both the linear and nonlinear regimes.

The absorption of surface acoustic waves (SAWs) with the frequencies $f$=30, 86, 140, 198, and 253~MHz and the changes in their speed are measured at magnetic fields up to 8 T and temperatures $T=1.7-15$K
in both the linear and nonlinear regimes. Here, SAWs are excited and detected by the interdigital transducers IDT1 and IDT2 created on the surface of a lithium niobate crystal. The sample under study is pinned down between these transducers by a spring. The propagation of a SAW (a Rayleigh wave) along the surface of lithium niobate ($U_{in}$ and $U_{out}$ are the input and output signals, respectively) is accompanied by the generation of an electric field penetrating into the sample and interacting with charge carriers in the conduction channel. The absorption of the SAW interacting with electrons and its phase change are measured as functions of the magnetic field, temperature, frequency, and intensity of this wave. Having the simultaneously measured absorption and phase change and using the formulas reported in~\cite{Dmitriev19}, it is possible to
determine the real and imaginary components of the complex AC conductivity, $\sigma_{xx}(\omega)\equiv \sigma_1-i\sigma_2$.

\begin{figure}[h!]
\centering
\includegraphics[width=8.5cm]{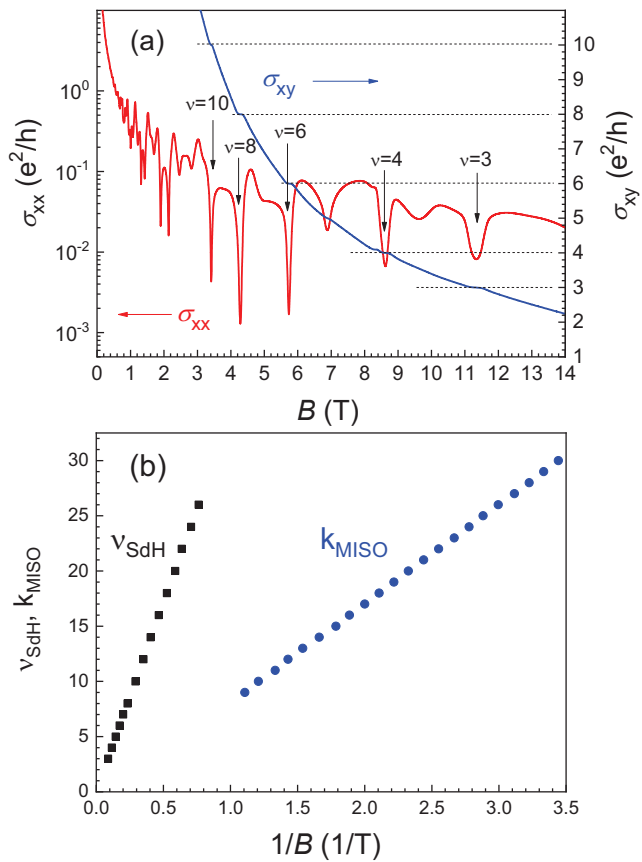}
\caption{(
(a) Magnetic field dependences of
the conductivities $\sigma_{xx}$ and $\sigma_{xy}$ at $T=2.65$~K. The filling factors are indicated above the oscillations. (b) Positions of maxima of magneto-intersubband oscillations
in $\sigma_{xx}$ ($k_{MISO}$) and of minima of SdH oscillations ($\nu_{SdH}$) versus the inverse magnetic field in the range of $0 \le B$ $\le 14$~T.
\label{fig2}}
\end{figure}

In the sample under study, the DC measurements of the $\rho_{xx}$ and $\rho_{xy}$ components of the magnetoresistance tensor are performed. These components depend on the temperature and electric current flowing across the sample. The magnetic field dependence of the conductivities $\sigma_{xx}$ and  $\sigma_{xy}$ at $T$=2.65~K (calculated using the measured components of the magneto-resistivity tensor by the formula $\sigma_{ik} =\rho_{ik}/(\rho_{xx}^2 +\rho_{xy}^2)$)) is shown in Fig.~\ref{fig2}.

The arrows in Fig.~\ref{fig2}a, drawn at the centers of the
$\sigma_{xy}$ plateau, correspond to $\nu=2\varepsilon_F/\hbar \omega_c$, where the
Fermi energy $\varepsilon_F$ is calculated for the total electron density $n_{\mathrm{tot}}$ in the quantum well at $B=0$ and $\omega_c$ is the cyclotron frequency. A factor of 2 is due to the spin splitting of the Landau levels. Figure~\ref{fig2} demonstrates a complicated pattern of oscillations, where SdH oscillations are observed at magnetic fields of 1–3 T (Fig.~\ref{fig2}a), the integer quantum Hall effect appears above 3 T, and intersubband oscillations manifest themselves at $B<1$~T (Fig.~\ref{fig2}b). Further, on, we discuss these issues in more detail.

\subsection*{Magnetic Fields $B>1$~T. Linear Regime}

In this sample, the complicated pattern of oscillations of the conductivity $\sigma_{xx}$ could be related to a certain filling factor $\nu$ by using the experimentally determined conductivities $\sigma_{xy}(B)$ at the plateau and their positions in the magnetic field at $T=2.65$~K. These values coincide with those calculated by the formula
$\nu=2\varepsilon_F/\hbar \omega_c$.

To calculate the Fermi energy as a function of the magnetic field for the system with a two-subband energy spectrum, we used the well-known expression for the electron density $n$:
\begin{equation} \label{fd}
n=\int \rho(\varepsilon) f_0 (\varepsilon)\, d\varepsilon.
\end{equation}
Here,
$\rho(\varepsilon)$ is the electron density of states and $f_0(\varepsilon)=\left[\exp \left(\frac{\varepsilon-\zeta}{k_BT}\right)+1\right]^{-1}$
is the Fermi–Dirac distribution function, where $k_B$ is the Boltzmann constant,
$\zeta$ is the chemical potential, and $\zeta_{T\to 0}$ is the Fermi
energy.

At nonzero applied magnetic field, neglecting the collisional broadening of Landau levels, the density of states can be written in the form
\begin{eqnarray}
\rho(\varepsilon)&=&\frac{eB}{2\pi \hbar c}\sum_{i=1,2} \sum_{s=\pm 1/2} \sum_{N=0}^\infty
\delta \left[ \varepsilon-\varepsilon_i \right. \nonumber \\ &&
\left.
-\hbar \omega_c (N+1/2)-sg\mu_0 B \right].
\end{eqnarray}
Here, $i$ is the number of the size quantization subband,
$\varepsilon_i$ is the energy corresponding to the bottom of the $i$th subband, $s = \pm 1/2$ is the spin projection on the magnetic field direction, $g$ is the electron spectroscopic splitting factor, and $\mu_0$ is the Bohr magneton. In a quantizing magnetic field, where $\hbar \omega_c \gg k_BT$, we can
make the replacement $f_0(\varepsilon) \to \Theta (\varepsilon_F -\varepsilon)$, after which the calculation of integral (\ref{fd}) becomes trivial. Each completely filled Landau level with a given spin projection makes the contribution $e/2\pi \hbar c$, and the Fermi level coincides with the upper, partially filled Landau level.

\begin{figure*}[t]
\centering
\includegraphics[width=14cm]{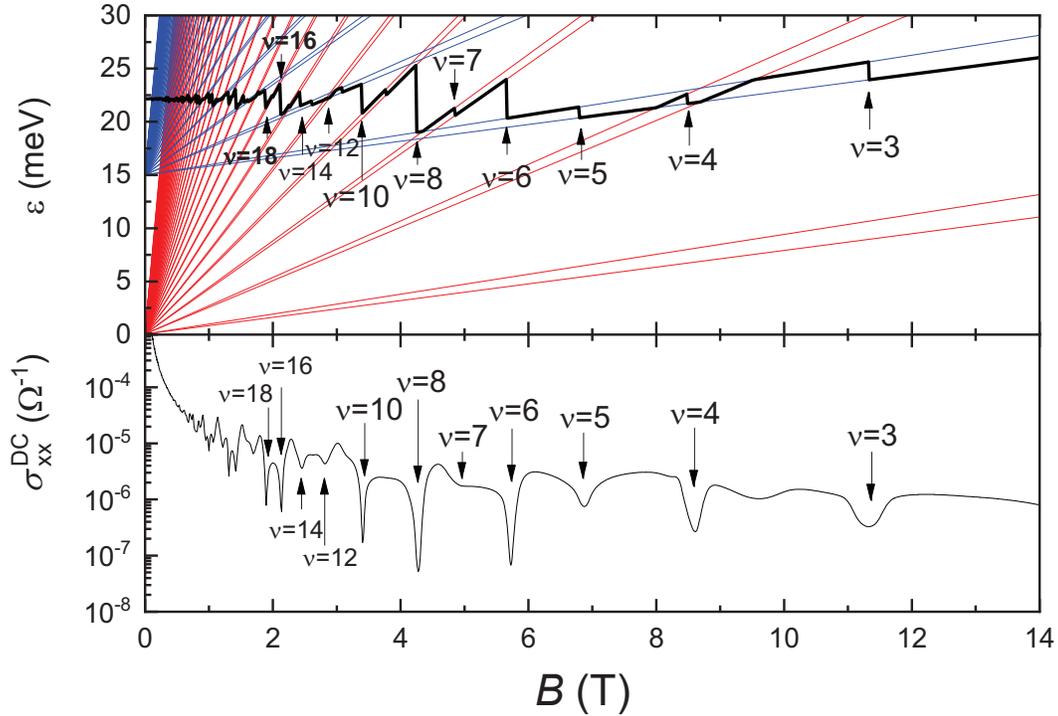}
\caption{
Fans of Landau levels for two subbands. Red lines denote Landau levels for the symmetric subband; these levels are spin split. Blue lines denote Landau levels for the antisymmetric subband; the $g$ factor is 1.3. The black line shows the magnetic field dependence of the Fermi level.
 \label{fig3}}
\end{figure*}

In Fig.~\ref{fig3}, we plot the “fans” of Landau levels for two subbands calculated using the following input data: the $S$ subband bottom $\varepsilon_1 \equiv 0$, the upper subband bottom $\varepsilon_2\equiv \Delta_{12} =15.5$~meV, the effective mass of electrons in GaAs $m^*=0.067 m_0$, and the electron spec- troscopic splitting factor $g$=1.3.

Using this energy diagram and Eq.~(\ref{fd}) for $T$=0 in the form
\begin{equation} \label{fd1}
n_{\mathrm{tot}}=\int_0^{\varepsilon_F} \rho(\varepsilon) \, d\varepsilon,
\end{equation}
we calculate the Fermi energy (see Fig.~\ref{fig3}) at the total charge carrier density $n_{\mathrm{tot}}=8.13\times 10^{11}$~cm$^{-2}$. Since the energy is measured from the bottom of the S subband, the Fermi energy at zero magnetic field is equal to the Fermi energy in the lower subband $\varepsilon_{F1}=22$~meV (which is proportional to the charge carrier density inthis subband). The calculated magnetic field dependence of the Fermi energy is shown in Fig.~\ref{fig3} by the black line.

The comparison of the top and bottom panels in Fig. 3 demonstrates that the positions of the minima in the oscillations along the magnetic field that are observed in the experiment and correspond to even occupation numbers (4, 6, 8, 10, …) are related to the Fermi level jumps between different subbands ($S$ and $AS$), whereas the positions of odd oscillations (5, 7) are related to the jumps between the spin-split Landau levels in each of the subbands. Thus, the above reasoning suggests that the complicated oscillation pattern of $\sigma_{xx}$ is due to jumps of the Fermi level between Landau levels of different subbands in the magnetic field.

The temperature dependence of electrical conductivity is studied using the contactless acoustic technique.
\begin{figure}[h]
\centering
\includegraphics[width=8cm]{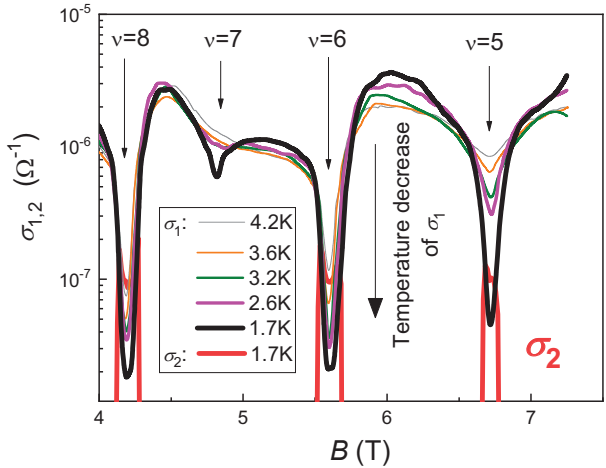}
\caption{Magnetic field dependences of the conductivity $\sigma_1$ at temperatures of 4.2, 3.7, 3.2, 2.7, and 1.7 K and the conductivity $\sigma_2$(B) at $T=1.7$~K; $f=86$~ MHz. The arrow indicates a decrease in temperature.
\label{fig4}}
\end{figure}

The magnetic field dependence of the linear AC conductivity at different temperatures is shown in Fig.~\ref{fig4}, where it is seen that the real component of the conductivity increases with the temperature in the regime of the quantum Hall effect. At $T=1.7$~K, $T=1.7$~ at the minima of conductivity, whereas $\sigma_2 \ll  \sigma_1$ between them because charge carriers at the minima of oscillations in the quantum Hall regime are localized, and the conductivity is determined by the hopping mechanism~\cite{Efros85}.

\subsection*{Magnetic Fields $B > 1$~T. The Range of the Integer Quantum Hall Effect.
Nonlinear Regime}

\begin{figure}[h]
\centering
\includegraphics[width=7.0cm]{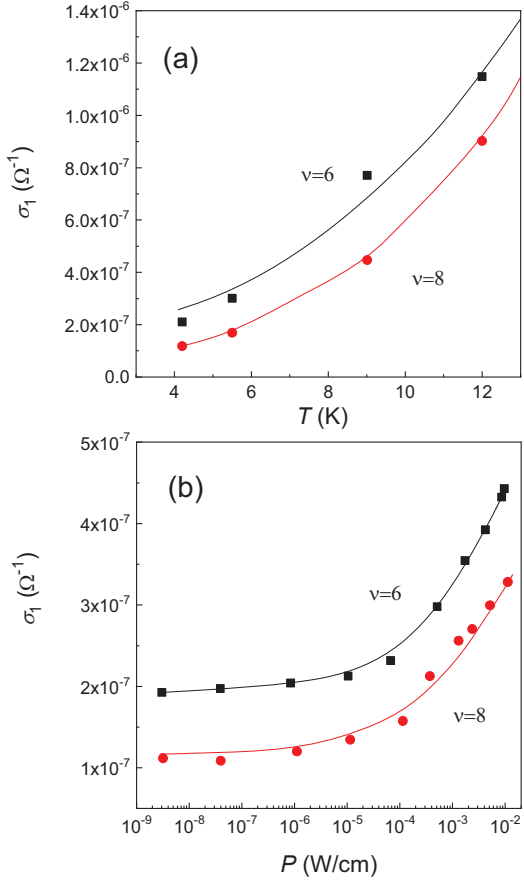}
\caption{(a) Temperature dependence of the
conductivity $\sigma_1$ at $\nu=6$  and 8. (b) Conductivity $\sigma_1$ versus the intensity $P$ (W/cm) of the surface acoustic wave at the
sample input; $T=4.2$~K; $f=30$~MHz.
\label{fig5}}
\end{figure}

Figure~\ref{fig5} shows the (a) temperature and (b) SAW intensity dependences of $\sigma_1$ at 30 MHz at the sample input corresponding to $\nu =6$ and 8, i.e., in the regime of the integer quantum Hall effect. As seen, $\sigma_1$ in this regime increases both with the temperature (Fig.~\ref{fig5}a) and with the SAW intensity (Fig.~\ref{fig5}b). Usually, this dependence of the conductivity on the SAW intensity is attributed to the heating of the electron gas by the electric field of the SAW. The estimate based on the comparison of Figs.~\ref{fig5}a and b shows that the electric field with an intensity of 0.01~W/cm heats the electron system being initially at 4.2~K only to about 7~K. The nonlinear effects in the DC conductivity arising in the regime of the integer quantum Hall effect were studied and analyzed in detail in a number of works (see, e.g., ~\cite{Ebert,Alexander-Webber}). It was found that nonlinearities are also due primarily to the heating of the electron gas by the electrostatic field.

The oscillation pattern of the conductivity $\sigma_1$ obtained by the DC measurements at magnetic fields up to 14~T in the linear regime is shown in Fig.~\ref{fig2}b. At low magnetic fields, the period of these oscillations is much larger than that of SdH oscillations. Since two size quantization levels exist below the Fermi level, these oscillations are assumingly intersubband oscillations. The plot of the positions of the maxima of these oscillations versus $1/B$ gives $\Delta_{12}=15.5$~meV, which coincides with the results of the Fourier analysis of magnetoresistance oscillations at $B<1$~T.

\subsection*{Magnetic Fields $B<1$ T}
\paragraph*{Linear regime.}

\begin{figure}[h!]
\centering
\includegraphics[height=6cm]{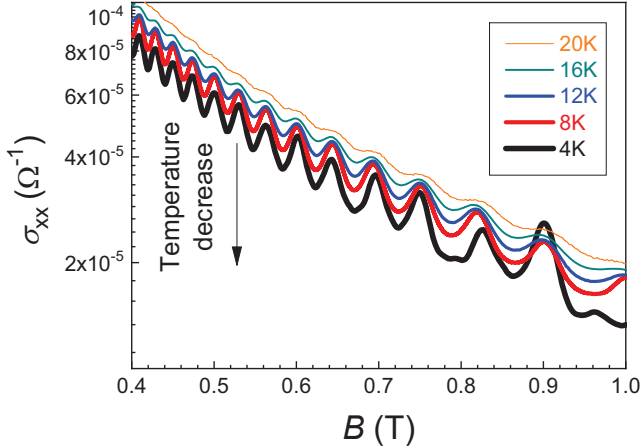} 
\caption{Magnetic field dependence of the DC conductivity $\sigma_{xx}$ in the regime of intersubband transitions at temperatures of 4, 8, 12, 16, and 20 K. The arrow
indicates a decrease in temperature.
	\label{fig6}}
\end{figure}

\begin{figure}[h!]
\centering
\includegraphics[width=0.48\textwidth]{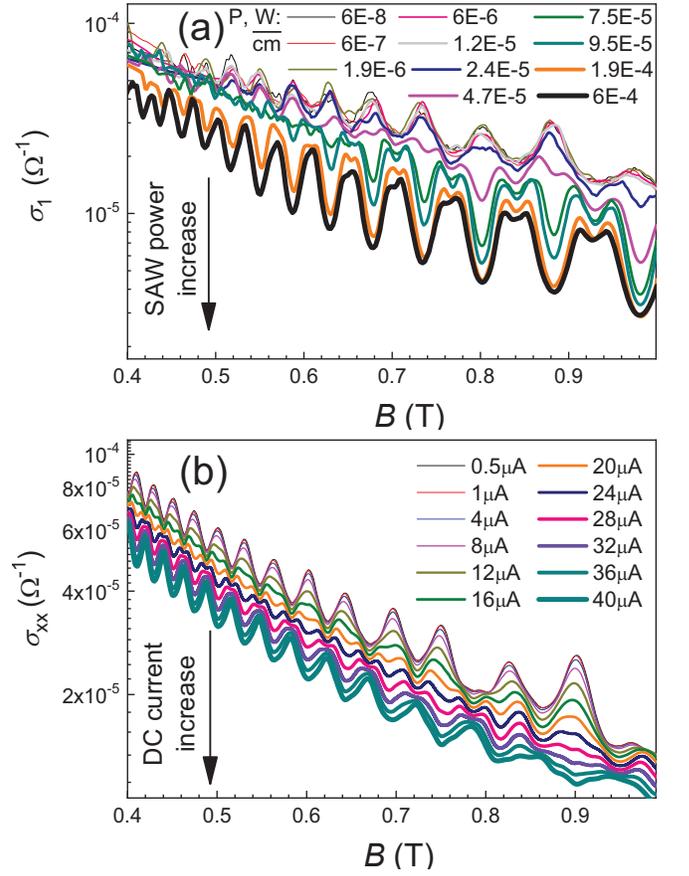}
\caption{Magnetic field dependence of the
conductivity $\sigma_1(B)$ at $T=4.2$~K and (a) at different SAW intensities at the sample input, $f=140$~MHz, and (b)
obtained by the DC measurements at different electric currents. The arrows indicate the increase in intensity and in the electric current at the sample input.
\label{fig7}}
\end{figure}

As mentioned in the Introduction, the conductivity in low magnetic fields is studied using DC measurements and acoustic spectroscopy. The magnetic field dependence of $\sigma_{xx}$ in the linear regime obtained by DC measurements at different temperatures is shown in Fig.~\ref{fig6}.
\paragraph*{Nonlinear regime.} The measured magnetic field dependence of the real part of $\sigma_{xx} (\omega)$ in the nonlinear regime is shown in Fig.~\ref{fig7}.
It is seen that different methods give a qualitatively similar behavior of the conductivity in the course of intersubband transitions in the nonlinear regime: with an increase in the electric current flowing across the sample or in the intensity of the acoustic wave, maxima and minima in the
conductivity alternate with each other. Comparing Figs.~\ref{fig6} and ~\ref{fig7}b, we can see that the temperature and DC current dependences of the conductivity have different forms. Namely, the conductivity increases slightly with the temperature, whereas the conductivity decreases with an increase in the electric field, and the conductivity maxima are replaced by minima with a further increase in $E$. This suggests that nonlinearity in intersubband transitions is hardly due to the heating of the electron gas, as in the regime of the quantum Hall effect.

\begin{figure}[h]
\centering
\includegraphics[width=8cm]{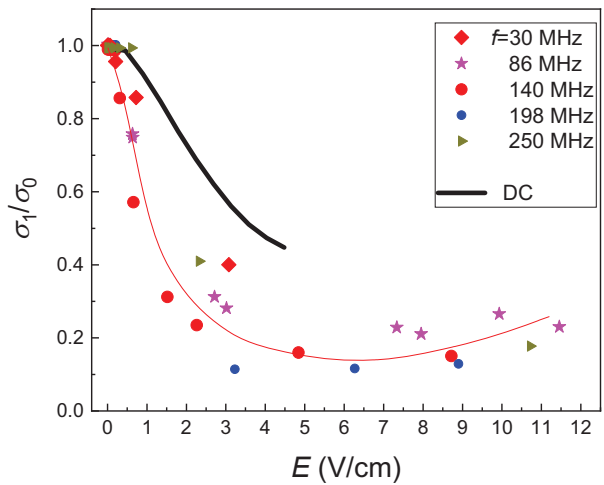}
\caption{
Electric field dependence of the ratio $\sigma_1/\sigma_0$ at different SAW frequencies (f) and in the DC regime at $B$=0.73~T ($k$=12 ), where $\sigma_0$ is the conductivity at the same frequency in the linear regime.
\label{fig8}}
\end{figure}

To compare the characteristics of nonlinear effects studied by different methods, we plot in Fig.~\ref{fig8} the dependence of the dimensionless conductivity, $\sigma_1/\sigma_0$, on the electric field $E$ applied across the sample at $k$=12 (at other $k$ values, the results are similar). In the acoustic measurements, the value of $E$ is determined by Eq.~(1) from ~\cite{Drichko97}. In the DC measurements, where the current $I_x$ flowing across the sample is varied, the $\mathbf{E}$ field has two components $E_x=\rho_{xx}I_x/d$  and $E_y=\rho_{yx}I_x/d$ such that $E_y \gg E_x$; i.e.,  $E_x$ is very small.

According to Fig.~\ref{fig8}, the dependence of $\sigma_1/\sigma_0$ on $E$ measured by the acoustic technique is independent of the SAW frequency within the measurement error and differs from the dependence of $\sigma_1/\sigma_0$ on $E_y$ obtained by the DC measurements. Moreover, at $E>6$~V/cm, the ratio $\sigma_1/\sigma_0$ measured by the acoustic technique begins to increase. This can be attributed to an increase in the temperature of the electron gas. The same effect is reported, e.g., in~\cite{Mamani09}, where DC measurements were performed with currents  exceeding
those used in our work by a factor of 7. Note that the
nonlinear behavior of $\sigma_1$ in the region of intersubband transitions is similar to that of the conductivity in balanced systems.

\section{Discussion of the results}

Let us summarize the main characteristic features of the revealed nonlinear effects in DC and AC conductivities. Unfortunately, there is currently no quan- titative theory of the nonlinear AC conductivity for two-subband unbalanced structures. For this reason, we give only a qualitative physical picture of nonlinear
effects in different magnetic field ranges.
\begin{itemize}
\item At $B > 3$~T, where the integer quantum Hall effect occurs, the electron heating by an electrostatic or high-frequency electric field induced by a propagating acoustic wave is responsible for the nonlinear behavior.
\item At $B < 1$~T, where magneto-intersubband oscillations occur in the linear regime, the nonlinear behavior of the conductivity is more diverse. First, in the DC case, the f lowing current generates an appreciable Hall field modulating the effective filling factor across the sample~\cite{Dietrich12}. This seems to be the main reason for the dependence of the nonlinear DC conductivity on the f lowing electric current, similar to that reported in ~\cite{Dietrich12}.

\item When the AC electric field is induced by the propagating acoustic wave, macroscopic Hall fields are absent because the $y$ components of the flowing currents have opposite directions in the regions corresponding to the neighboring half-periods of the SAW. As a result, the average $y$ component of the current (and hence, the macroscopic Hall field) vanishes.
\end{itemize}
In such a situation, the nonlinear behavior can apparently be attributed to the so-called \textit{quantal heating}~\cite{Dmitriev05}. Just this interpretation of the results is adopted in several experimental studies~\cite{Bykov08,Mamani09,Zhang07,Zhang09}. This mechanism is due to the quantization of the
electron spectrum in the magnetic field. As a result, the energy dependence of the electron density of states has the form of a set of narrow peaks. A change in the relative positions of the peaks in the density of states corresponding to different subbands in the applied magnetic field leads to a magnetic field dependence of the probabilities of intersubband transitions. This gives rise to the oscillations of the conductivity.

The probabilities of intersubband transitions depend both on the mutual arrangement of the peaks in the density of states (coinciding with the Landau levels) and on the differences in the filling factors of these states. With an increase in the electric field, the energy distribution of electrons becomes nonequilibrium. This distribution function is then determined by the equation of diffusion, and the diffusion coefficient for energy is proportional to the electric field squared. Therefore, this is the so-called spectral diffusion, which leads to a decrease in the difference between the occupation numbers of the initial and final states. The quantitative analysis~\cite{Mamani09} of the nonlinear DC conductivity shows that the quantization of the spectrum and the nonequilibrium of the distribution function make contributions to the conductivity with opposite signs. For this reason, with an increase in the electric field, the maxima in the magneto-oscillation pattern are transformed to minima.

A detailed interpretation of the observed phenomena requires a quantitative nonlinear theory of the AC conductivity of a two-subband electron system in the applied magnetic field. In such a theory, it is necessary to take into account the Landau quantization, elastic and inelastic scattering of electrons by each other, structural defects, and phonons, as well as the acceleration of electrons by the applied electric field. As mentioned above, a detailed analysis of the DC case was performed in~\cite{Dmitriev05}. We believe that the experimental results obtained in this work should stimulate the development of such a theory for the AC conductivity.

\section*{Conclusion}

In this work, the contactless acoustic technique has been applied for the first time to study linear and non-linear AC conductivities in an n-GaAs/AlAs heterostructure that has two occupied size quantization levels (with different carrier densities) and, therefore, a two-subband energy spectrum. It is shown that the nonlinear behavior of the AC conductivity in two-subband heterostructures differs markedly from that characteristic of usual heterostructures with a single occupied size quantization level.

In conventional heterostructures, the linear AC conductivity in the regime of SdH oscillations and in the frequency range under study is independent of the SAW frequency and coincides with the DC conductivity. With an increase in the temperature, SAW intensity, or current, these oscillations are suppressed because of the heating of the electron gas.

In two-subband heterostructures, the linear AC and DC conductivities are also close to each other. At the same time, the nonlinear behaviors of conductivities are significantly different. Thus, the study of the nonlinear AC conductivity provides additional information on the magnetoconductivity of the quasi-two-dimensional electron gas.

We believe that the significant difference in the behavior of nonlinear AC and DC conductivities, which is the main result of this work, suggests an important role of the macroscopic Hall field. Such field is generated in the DC case and is absent in the AC one. Note once again that a detailed interpretation of experimental results reported in this work requires significant progress in the development of the quantitative theory of the nonlinear AC conductivity.

\section*{Funding}
This work was supported by the Russian Foundation for Basic Research (project nos. 19-02-00124 and 20-02- 00309) and by the Presidium of the Russian Academy of Sciences.

\vfill\eject

\end{document}